\documentclass[aps,reprint,prb]{revtex4-1}
\usepackage{amsmath}
\usepackage{graphicx}
\usepackage{dcolumn}
\usepackage{bm}

\begin{document}

\preprint{APS/123-QED}

\title{Efros-Shklovskii variable range hopping and nonlinear transport in 1T/1T$^{\prime}$-MoS$_{2}$} 
\author{Nikos Papadopoulos}
\email{Email: n.papadopoulos@tudelft.nl\\}
\author{Gary A. Steele}
\author{Herre S. J. van der Zant}

\affiliation {\small \textit Kavli Institute of Nanoscience, Delft University of Technology, Lorentzweg 1, Delft 2628 CJ, The Netherlands\\}



\begin{abstract}
\noindent
We have studied temperature- and electric-field dependent carrier transport in single flakes of MoS$_{2}$ treated with n-butyllithium. The temperature dependence of the four-terminal resistance follows the Efros-Shklovskii variable range hopping conduction mechanism. From measurements in the Ohmic and non-Ohmic regime, we estimate the localization length and the average hopping length of the carriers, as well as the effective dielectric constant. Furthermore, comparison between two- and four-probe measurements yield a contact resistance that increases significantly with decreasing temperature. 

\end{abstract}

\maketitle

\section{\label{sec:level1} Introduction}

\noindent
Transition metal dichalcogenides (TMDCs) form a family of van der Waals crystals with the general formula MX$_2$, where M is a transition metal and X a chalcogen atom.  Molybdenum disulphide is the most known among the TMDCs and in its natural form (2H phase) it is a layered semiconductor with a band gap of 1.3 eV in bulk and 1.8 eV in monolayers.\cite{mak_atomically_2010} 2H-MoS$_{2}$ has attracted a lot of interest due to its use in field-effect transistors (FETs),\cite{radisavljevic_single-layer_2011} photodetectors,\cite{lopez-sanchez_ultrasensitive_2013} and its rich spin-valley physics.\citep{cao_valley-selective_2012} Unlike 2H-MoS$_{2}$, the 1T-MoS$_{2}$ phase has metallic properties and an octahedral stucture.\cite{wypych_1t-mos_1992} This phase is metastable and relaxes to the distorted 1T$^{\prime}$ one with clustering of the Mo sites and formation of Mo chains.\citep{sandoval_raman_1991}$^{,}$\cite{qin_real-space_1991} The 1T$^{\prime}$ phase is semiconducting whose band gap has not been measured directly, but calculations  yield values ranging from 0.08 eV\citep{qian_quantum_2014} to 0.8 eV.\citep{hu_new_2013}

High doping levels can cause a phase transition from the 2H to the 1T and 1T$^{\prime}$ phases, which can be achieved chemically via charge transfer through intercalation of alkali metals,\citep{imanishi_study_1992} by exposure to electron beam irradiation \citep{katagiri_gate-tunable_2016} or by metallic adatom adsorption on the surface.\citep{enyashin_new_2011} This phase transition has been studied extensively and  was found to take place with gliding of the sulfur atom planes.\cite{lin_atomic_2014} Unfortunately, the above processes convert the 2H phase to the 1T and 1T$^{\prime}$ phases (1T/1T$^{\prime}$), but they also leave some domains of the semiconducting 2H phase inside the MoS$_{2}$ lattice.\citep{eda_coherent_2012} Nonetheless, the resulting sheets have very different electronic and chemical properties than the natural 2H-MoS$_{2}$.

Although there is a large variety of studies on 1T/1T$^{\prime}$-MoS$_{2}$ and related heterostructures, there are not so many investigations on their electrical properties. Recently, temperature dependent two-terminal transport measurements showed that electrons in 1T/1T$^{\prime}$-MoS$_{2}$ from chemical treatment are localized inside the metallic patches of the 1T phase, leading to Mott variable-range-hoping (VRH).\cite{kim_electrical_2016} Here, we report on four and two-probe measurements on few-layer 1T/1T$^{\prime}$-MoS$_{2}$ flakes, obtained from a n-butyllithium treatment. We find that the channel resistance increases dramatically as the temperature decreases. Comparison between the two measurement configurations yields a small contact resistance at room temperature that increases considerably at low temperatures. We find that the temperature dependence of the resistance obtained with four terminal-measurements in the Ohmic regime fits the Efros-Shklovskii VRH mechanism better than the Mott-VRH model. Furthermore, we study the nonlinear transport at low temperatures with two-probe measurements. While at low bias (Ohmic-regime) the temperature dependence of the resistance is strong, at high electric fields this dependence is suppressed and the device operates in the non-Ohmic and electric-field activated regime.\cite{tremblay_activationless_1989}

\section{\label{sec:level1} Results}

\subsection{\label{sec:level2} Phase transition and fabrication}

\noindent
Thin MoS$_{2}$ flakes were obtained using the scotch tape technique and transferred on 285 nm SiO$_{2}$/Si substrates via a PDMS dry transfer method.\cite{castellanos-gomez_deterministic_2014} A transferred flake before the chemical treatment is shown in Fig. 1a. The flakes were immersed in n-butyllithium (1.6 M in hexane) for more than 48 hours and after extraction, the substrates were washed with hexane and deionized water to remove excess lithium. After the chemical treatment, there is a color change of the flake as can be seen in Fig. 1b. Another way to verify the phase transition is with Raman spectroscopy. Figure 1d depicts the spectrum of a flake after extraction from the n-butyllithium solution (black line). The $J_{1}$, $J_{2}$ and $J_{3}$ peaks that originate from the 1T$^{\prime}$ phase (green labels), the $A_{1g}$ peak from the 1T phase, as well as the $E_g$, $E_{2g}^{1}$ and $A_{1g}$ peaks from remnant patches of the initial 2H phase can be seen.\cite{sandoval_raman_1991}$^,$\citep{hu_new_2013} The Raman spectrum therefore indicates that the phase transition was incomplete, in line with previous reports. \cite{eda_coherent_2012}$^,$\cite{voiry_conducting_2013}

\begin{figure*}
  \begin{center}
    \includegraphics[width=15.237cm]{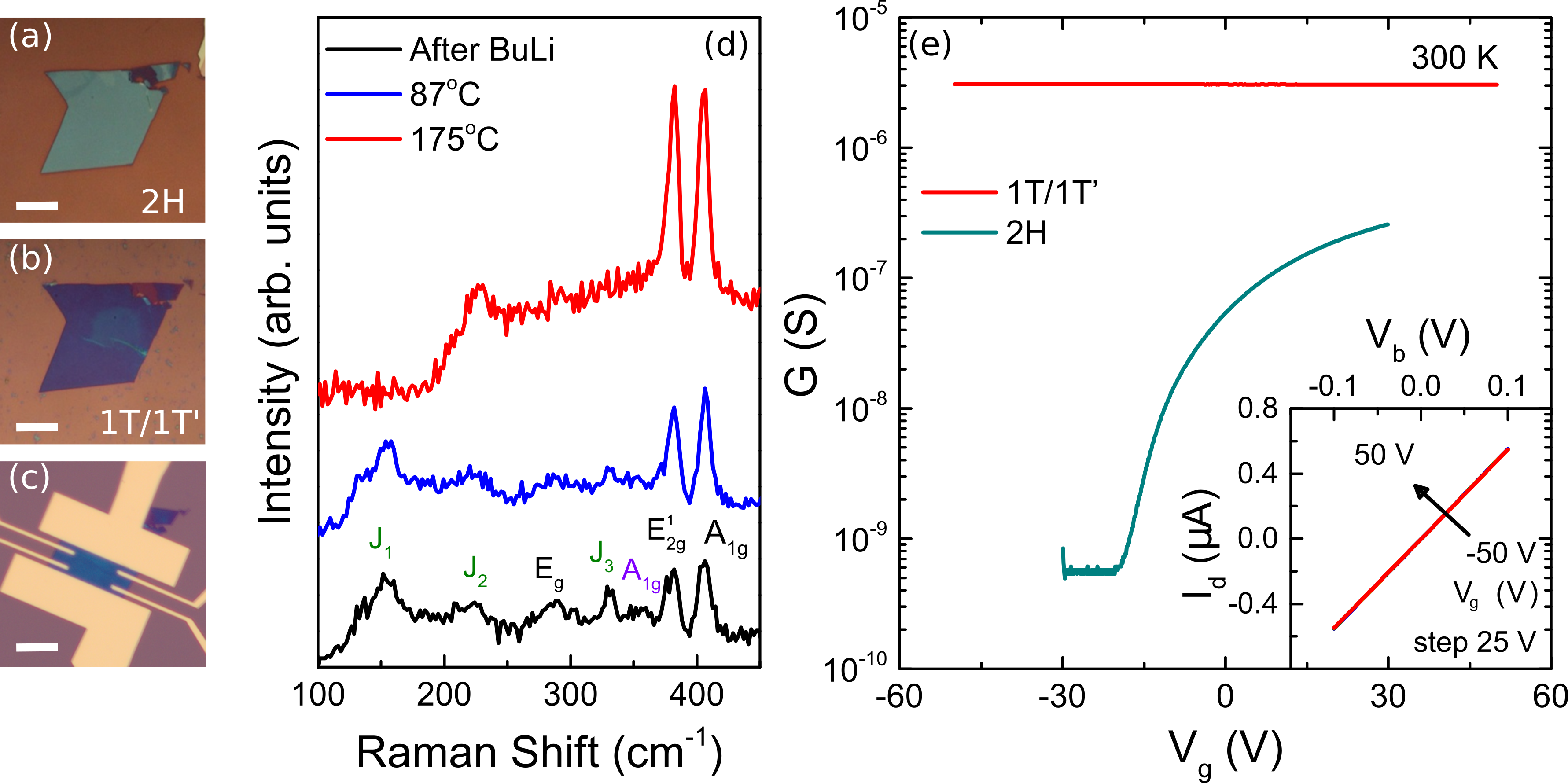}
  \end{center}
  
  \caption{\small Room-temperature characterization of 1T/1T$^{\prime}$-MoS$_2$ flakes and devices. Optical images of an MoS$_{2}$ flake after exfoliation (a), after immersion in n-butyllithium (b) and of a fabricated device (c). Scale bar is 15 $\mu$m. (d) Raman spectra from 1T/1T$^{\prime}$-MoS$_{2}$ devices prepared under different conditions: Unbaked (black line), baked at 87 $^o$C (blue line) and baked at 175 $^o$C (red line). (e) Two-terminal conductance as a function of the back-gate voltage for 2H-MoS$_{2}$ (grey curve) and 1T/1T$^{\prime}$-MoS$_{2}$ (red curve). The inset shows the transfer characteristics of the 1T/1T$^{\prime}$ channel, for five back-gate voltages from -50 V to 50 V with a step of 25 V; all curves fall on top of each other.}
\end{figure*}

After inducing the phase transformation, we proceed to device fabrication, for which standard e-beam lithography was used with a single layer PMMA resist. To preserve the 1T/1T$^{\prime}$ phases, heating the PMMA above 95 $^o$C must be avoided, which is the temperature at which the 1T/1T$^{\prime}$ to 2H phase transition is expected to take place.\cite{wypych_1t-mos_1992} We have found that baking the PMMA resists at 87 $^o$C in a vacuum oven for a couple of minutes is sufficient to preserve the 1T/1T$^{\prime}$ phase. This can be seen in Fig. 1d, where the Raman spectrum before (black curve) and after baking (blue curve) is similar, indicating that there is no substantial composition change of the flake. In contrast, flakes with PMMA baked at 175 $^o$C for 3 minutes show a significant reduction in the intensities of the $J_{1}$, $J_{2}$ and $J_{3}$ peaks with a change in the background and an increase in the peak intensity from the 2H phase. 

After PMMA patterning via e-beam lithography, e-beam metal evaporation was used to evaporate 5 nm of Ti and 70 nm of Au to form the contacts. Several Hall bars and other multi-terminal devices for transport measerements have been fabricated. The advantage of n-butyllithium treatment and post-fabrication, compared to a treatment after device fabrication, is that below the contacts there is 1T/1T$^{\prime}$-MoS$_{2}$, which can provide better Ohmic contacts according to earlier reports.\cite{kappera_phase-engineered_2014}$^,$\cite{kappera_metallic_2014} One of the final devices is shown in Fig. 1c.

Figure 1e shows a room-temperature, two-probe electrical characterization of devices with 2H (grey) and 1T/1T$^{\prime}$-MoS$_{2}$ (red) flakes. The $G$-$V_{g}$ curves were obtained by applying a DC voltage bias between source and drain and measuring the source-drain current while sweeping the back-gate voltage. In the case of the sample with the 1T/1T$^{\prime}$-MoS$_{2}$, the back-gate modulation of the conductance is negligible, while in the untreated 2H-MoS$_{2}$ sample the curve is semiconducting n-type with a high ON/OFF ratio. The zero transconductance in the case of the treated sample shows that the Fermi level lies inside the conduction band and that the material has a high electron density.  The inset shows current-voltage characteristics from the device with the 1T/1T$^{\prime}$-MoS$_{2}$ channel at different back-gate voltages; these are linear, indicating Ohmic behavior.

\subsection{\label{sec:level2} Temperature dependence in the Ohmic regime}

\noindent
To investigate the electrical properties of thin 1T/1T$^{\prime}$-MoS$_{2}$ flakes, we studied their four- and two-terminal resistance as a function of temperature in two devices. Figure 2a shows current-voltage curves of two-terminal measurements that remain linear (Ohmic) down to liquid nitrogen temperatures. The decreasing slope indicates that the resistance increases upon cooling. Using a four-probe configuration, we extract the resistance of the channel in the Ohmic regime as a function of temperature by applying currents of $\pm$100 nA between source and drain, while measuring the voltage drop across the channel. Figure 2b shows the two-probe and four-probe resistance as a function of temperature; both exhibit a strong temperature dependence displaying semiconductor-like behavior. The four-terminal resistance increases from 5 k$\Omega$ at room temperature to 180 k$\Omega$ at 90 K, while the two-probe resistance reaches 700 k$\Omega$ at 90 K. From this data, it is clear that although the 1T/1T$^{\prime}$ state shows a reduced resistance at room temperature and no gate voltage dependence, at low temperatures it exhibits an insulating state.

From the data in Fig. 2b the contact resistance of the device can be estimated from the formula $R_{c}=0.5(R_{2T}-(l_{2T}/l_{4T})R_{4T})$, where $R_{2T}$ is the two-terminal and  $R_{4T}$ the four-terminal resistance, $l_{4T}$ the length between the voltage probes and where $l_{2T}$ is the length between the current contacts. At 275 K $R_{c}$ is around 5.2 k$\Omega$ and increases considerably with decreasing temperature, reaching 70 k$\Omega$ at 90 K (Fig. 2c). The contact resistance in the two devices was found to be less than 20\% of the total resistance between 90 K and 275 K.

To probe the nature of this localization, we analyze the temperature dependence of the four-terminal conductance of the device. The increase in the resistance indicates that the carrier transport takes place via hopping processes of the localized carriers. There are several models for hopping transport in solids. In the nearest neighbor hopping model (NNH) the resistivity is proportional to exp$(E_{A}/k_{B}T)$, with $E_{A}$ the activation energy.\cite{gantmakher_electrons_2005} The general form for VRH assisted transport on the other hand is $\rho\propto$ exp$((T_{0}/T)^{a})$, where $T_{0}$ is a characteristic temperature. For two-dimensions and in the case of Mott-VRH, the exponent $a$ is equal to $1/3$ and the electrons hop between states that are spatially further apart but closer energetically.\cite{mott_conduction_1969} In the case of Efros-Shklovskii (ES) VRH the exponent $a = 1/2$ and hopping takes place under the influence of strong electron-electron interactions.\cite{efros_coulomb_1975}

\begin{figure*}
  \begin{center}
    \includegraphics[width=13.25cm]{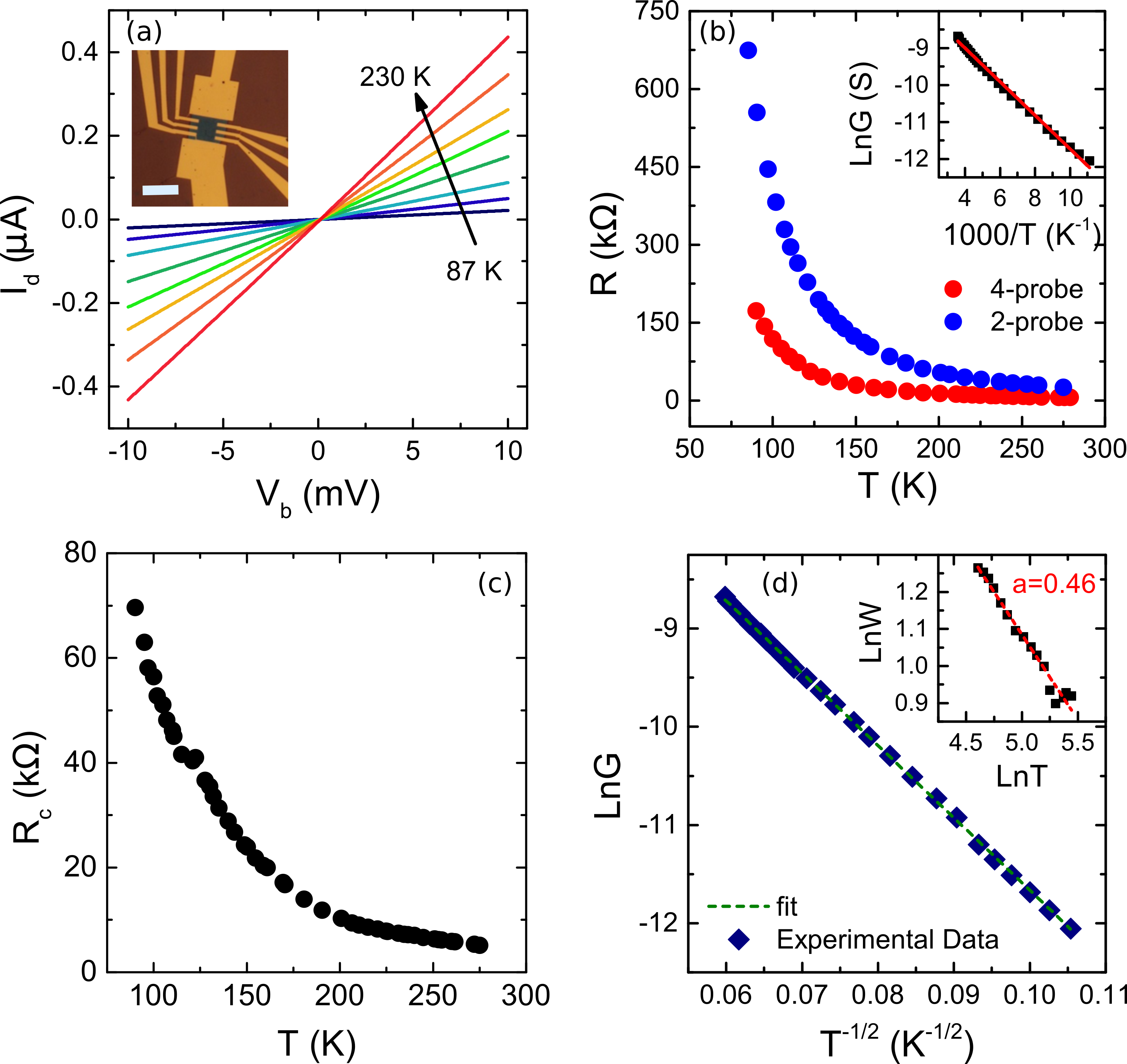}
  \end{center}
  \caption{\small Four-terminal transport measurements from room-temperature to liquid nitrogen temperatures to determine transport mechanisms. (a) Current-voltage ($I_d$-$V_b$) characteristics from 87 K to 230 K. In this temperature range, the observed low-bias $I_d$-$V_b$ curves are linear with resistances below 1 M$\Omega$. Inset shows an optical image of the device that was used. Scale bar is 7 $\mu$m. (b) Two-terminal (blue) and four-terminal (red) resistance as a function of temperature. Inset shows the dependence of natural logarithm of the four-terminal conductance from the inverse of the temperature 1000$/T$. The deviation of the data from a linear relation (red line) indicates that the transport mechanism is not nearest neighbor hopping. (c) Contact resistance as a function of temperature: at low temperatures, the contact resistance diverges. (d) Ln$G$ as a function of $T^{-1/2}$. The symbols are experimental data and the green dashed line the linear fit to them. Inset shows Ln$W$ (see main text for definition) as a function of Ln$T$. The linear fit yields an exponent $a = 0.46 \pm 0.02$, consistent with the ES-VRH model. }
\end{figure*}

As it can be seen from the inset of Fig. 2b, which depicts the nastural logarithm of the four-terminal  conductance Ln$G$ as a function of 1000$/T$, the data do not follow a straight line so that the NNH model cannot explain the conduction mechanism of 1T/1T$^{\prime}$-MoS$_{2}$. To analyze this further, one can plot Ln$G$ vs.\ $T^{-1/2}$; in the case that the transport is governed by ES-VRH, the data should show a linear relation. The data in Fig. 2d shows such a plot, and the linear relation indeed suggests an ES-VRH mechanism. To confirm the exponent, one can also plot Ln$W$ as a function of Ln$T$, where $W=-\partial$Ln$\rho/\partial$Ln$T\propto{a{(T_{o}/T)}^{a}}$. The slope of Ln$W$ vs.\ Ln$T$ is equal to the exponent $a$. Such a plot is shown in the inset of Fig. 2d. From a linear fit, we extract an exponent $a = 0.46 \pm 0.02$, close to the exponent expected for an ES-VRH mechanism.\cite{joung_efros-shklovskii_2012}$^,$\cite{liu_variable_2016} Similarly from a second device (device B) $a=0.43\pm0.02$. Interestingly, if we do the same analysis for two-terminal measurements that include the contact resistance, we find a slope for device A of 0.31$\pm$0.02 and for device B of 0.49$\pm$0.03, highlighting the importance of four-terminal measurements for the determination of such exponents.

Returning to the analysis of Ln$G$ vs.\ $T^{-a}$, we can also compare different linear fits taken for different exponents for the temperature on the x-axis.  In the plot of Ln$G$ vs.\ $T^{-1/2}$, a linear fit yields a residual sum of squares error of 0.0076. Similarly, in a plot of Ln$G$ vs.\ $T^{-1/3}$ (not shown), the linear fit yields a value of the residual sum of squares error equal to 0.025. The smaller value of the residuals in the former case verifies that the Efros-Shkolvskii mechanism explains better our results than the Mott-VRH model. From this analysis, we can also extract the slope in a Ln$G$ vs.\ $T^{-1/2}$ plot, from which the characteristic temperature of the ES hopping ($T_{ES}$) can be determined. For the two devices, we find a $T_{ES}$ of 5426 K (device A) and 7898 K (device B).

\begin{figure}
  \begin{center}
    \includegraphics[width=6.333cm]{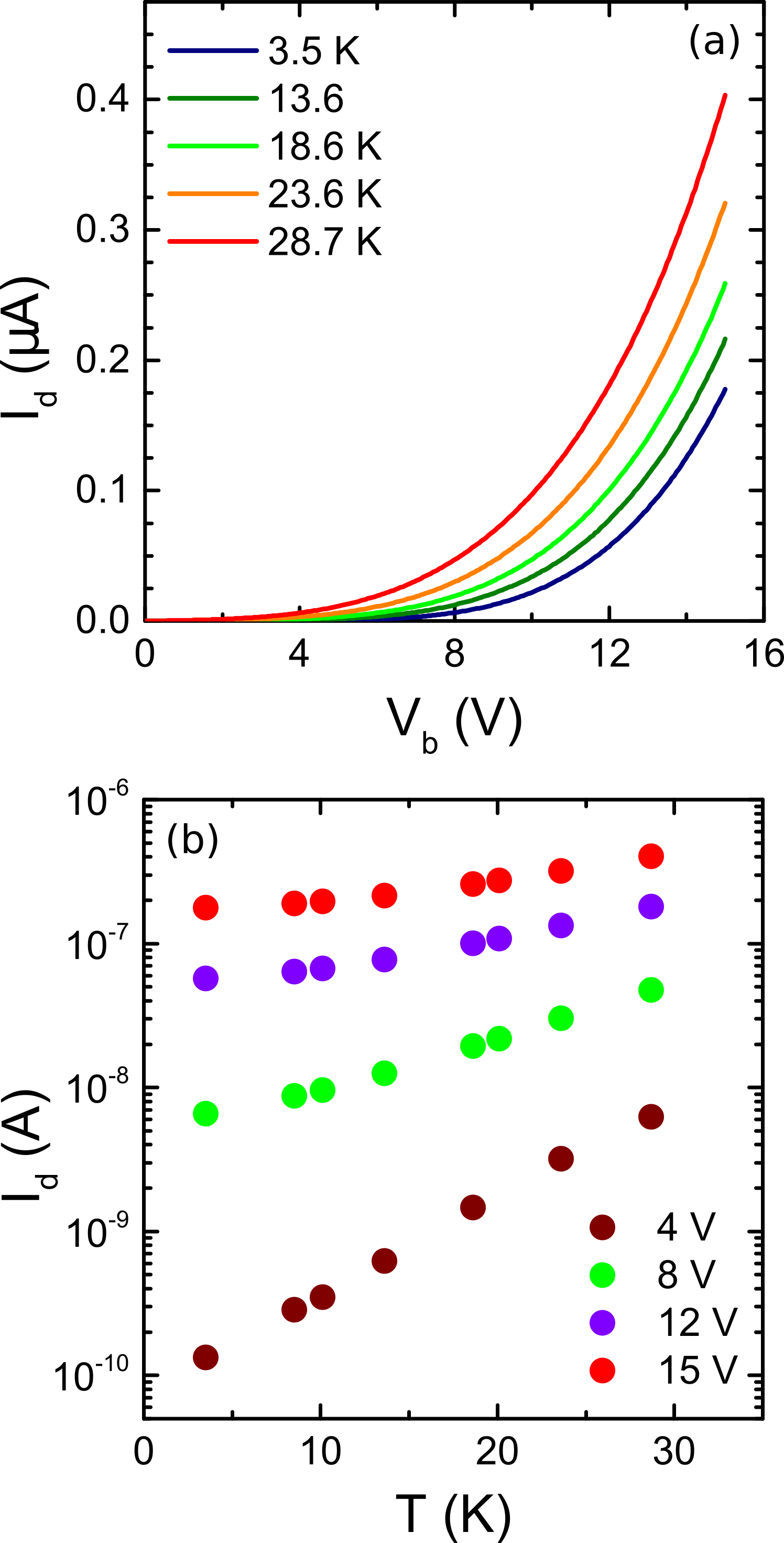}
  \end{center}
  \caption{\small Two-terminal nonlinear transport at low temperatures. (a) Nonlinear current-voltage characteristics at low temperatures. (b) Semi-logarithmic plot of the current $I_d$ as a function of temperature for different bias voltages $V_b$. The crossover from a strong temperature dependence to a weak temperature dependence of the channel current is more clearly seen.}
\end{figure}

\subsection{\label{sec:level2} Electric-field dependence in the non-Ohmic regime}

\noindent
Another aspect of hopping conduction is the field-assisted motion of charge carriers between localized states.\cite{shklovskii_hopping_1973} This field-assisted hopping leads to nonlinear transport characteristics and above a critical electric field the conductivity becomes temperature independent. According to the ES-VRH model, the dependence of the resistivity from the electric field ($E$) is given by:\cite{tremblay_activationless_1989}$^,$\cite{a._v._dvurechenskii_activationless_1988}$^,$\citep{yu_variable_2004}
\begin{equation}
  \rho\propto{exp{({E_{ES}}/{E})^{1/2}}},
  \label{eq:E3}
\end{equation}  
where $E_{ES}$ is a characteristic field connected to the localization length($\xi$) and to $T_{ES}$ by the relationship: $E_{ES}=k_{B}T_{ES}/e\xi$.\cite{a._v._dvurechenskii_activationless_1988}

Bellow a critical field $E_c(T)$ transport follows an Ohmic dependence and is in the strongly temperature-dependent regime, since the phonons assist the hopping processes. Above $E_c(T)$ the carriers have enough energy to pass the Coulomb barrier and the temperature dependence is suppressed. As $E_c(T)$ is temperature dependent and decreases as temperature is lowered, nonlinear current-voltage curves are therefore more prominent at low temperatures; furthermore, at low temperatures the channel resistance can become very high and four-terminal measurements are therefore more challenging for studying the electric-field dependence of transport.

In Fig. 3a we plot two-terminal $I_d$-$V_b$ curves of the device (device B in this case), at temperatures between 3.5 and 30 K; they are highly nonlinear. At a temperature of 3.5 K and with a bias of 4 V the resistance is 4.4 G$\Omega$, while for 15 V it declines to 150 M$\Omega$. Despite the high bias we did not observe  electrical breakdown of the devices. Note, that the large channel and contact resistances (G$\Omega$), do not allow to perform four-terminal measurements due to the internal resistance of the voltmeter and the voltage limits of our isolation amplifiers. Nonlinear transport characteristics in the current-voltage curves in this device were also observed in four-terminal measurements at temperatures between 85 K and 105 K (see Fig. S1), but the activationless regime is not accessible due to the high critical field ($E_c$) at these temperatures.

The crossover from strong to weak temperature dependence can be seen more clearly in Fig. 3b, which depicts a semi-logarithmic plot of the current as a function of temperature for different  bias voltages ($V_b$). The plot indicates that the suppression of the temperature dependence takes place above 12 V. This translates to an electric field of 2$\times$10$^6$ V/m for a channel of 6 $\mu$m. At $V_{b}$=4 V the ratio between the current at 3.5 K and 30 K is on the order of 50, while for 15 V this ratio is about 2.

Assuming that the nonlinearity in the current-voltage curves arises from ES-VHR with a negligible contribution from the contacts, we can then extract the ES electric field from the non-Ohmic regime. The value of $E_{ES}$ can be obtained by plotting the Ln$I_d$ as a function of $E^{-1/2}$ as depicted in Fig. 4 for different temperatures. As the electric field increases (left hand side of the plot) the curves from different temperatures converge to a single line at the temperature dependent critical field ($E_c(T)$). Equation \eqref{eq:E3} indicates that a  least square fit to the linear part of the Ln$I_d$-$E^{-1/2}$ curve provides an estimate for $E_{ES}$. For the lowest temperature (3.5 K), which fulfills this condition, we obtain a slope of 14347 (V/m)$^{1/2}$ that corresponds to an electric field of 2.06$\times$10$^8$ V/m (inset of Fig. 4). For the other device, $E_{ES}$ is found to be 3.8$\times$10$^8$ V/m.

\begin{figure}
  \begin{center}
    \includegraphics[width=7.5cm]{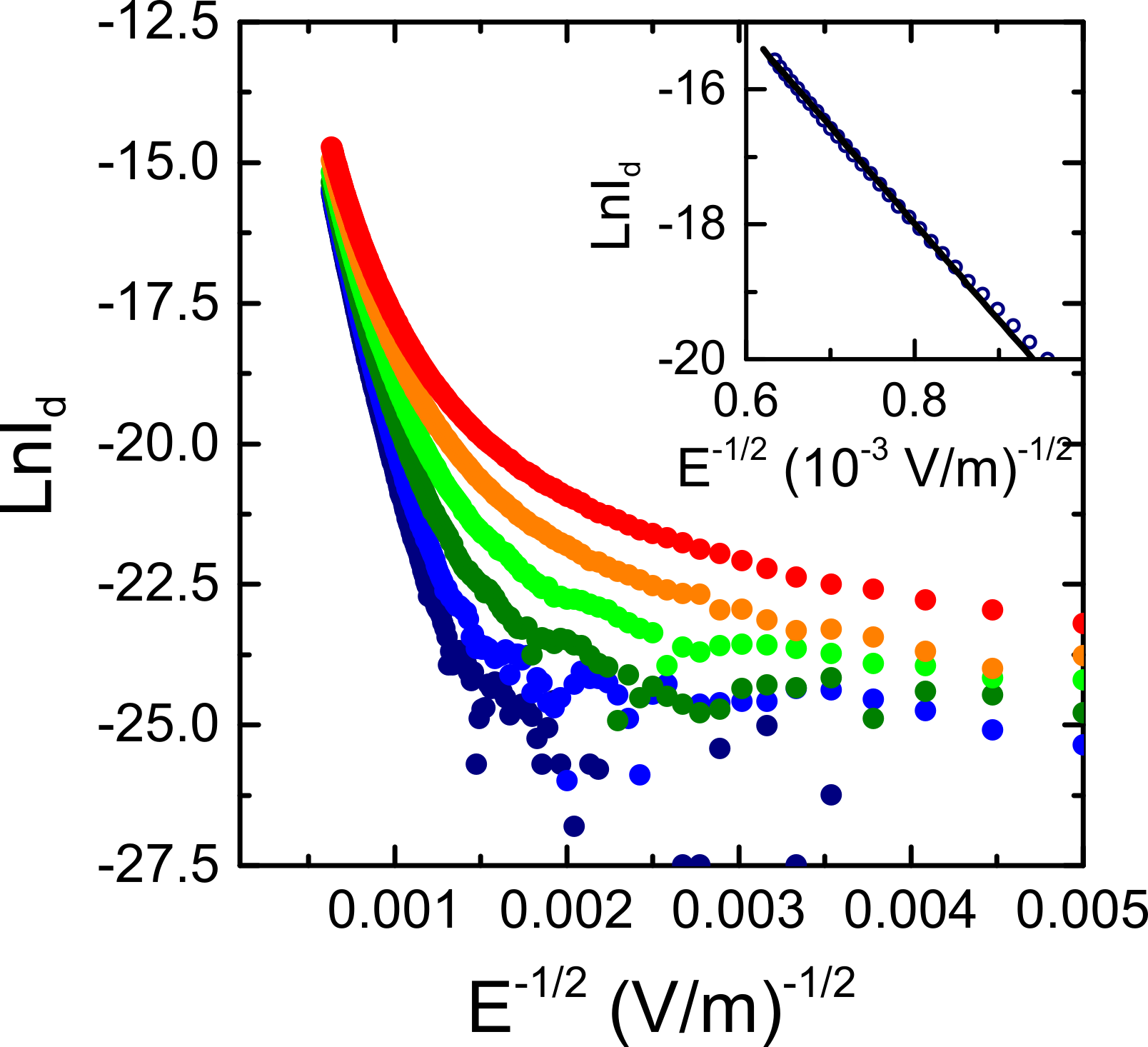}
  \end{center}
  \caption{\small Determination of the characteristic field $E_{ES}$ from the nonlinear transport data. Ln$I_d$ as a function of $E^{-1/2}$ for temperatures of 3.5 K (navy), 8.6 K (blue), 13.6 K (dark green), 18.6 K (light green), 23.6 K (orange) and 28.7 K (red). The slope of the linear part at high electric fields and at 3.5 K yields the parameter $E_{ES}^{1/2}$ (inset).}
\end{figure}

\section{\label{sec:level1} Discussion}
\noindent
From the values of $E_{ES}$ and $T_{ES}$ determined from the experiment, the localization length $\xi$ can be estimated using the relationship $\xi=k_{B}T_{ES}/eE_{ES}$. For device A the localization length is 1.2 nm, while for device B the localization length is 3.3 nm. The average hopping length, which for ES-VRH is given by $\overline{r}=\xi(T_{ES}/T)^{1/2})$, can be estimated using the experimental values of $\xi$ and $T_{ES}$. At 300 K, the average hopping distances of the carriers in Device A and Device B are 5 and 17 nm, respectively. The obtained values of $\overline{r}$ favor the physical picture of electron hopping from one 1T phase patch to another as previously suggested by Kim et al.\cite{kim_electrical_2016} 

Another parameter that can be calculated from the data is the effective dielectric constant of Li treated MoS$_2$, which in this case originates from the 1T$^{\prime}$ and 2H phases between the 1T phase domains. According to the ES-VRH model the critical temperature is given by:\cite{Shklovskii_Doped_Semiconductors} 

\begin{equation}
  T_{ES}=\frac{2.8e^2}{4\pi\epsilon_{0}\epsilon_{r}k_B\xi},
  \label{eq:E4}
\end{equation}
\noindent
where e is the electron charge, $\epsilon_{r}$ is the dielectric constant and $\epsilon_{0}$ the electric permittivity of the vacuum. We estimate effective dielectric constants of 7 and 2 for the thick and thin flakes, in line with the order of magnitude expected for MoS$_2$.\cite{li_charge_2016} We note that this estimation of the dielectric constant does not take into account the possibly metallic character of the material, which could change the screening characteristics. To get better estimates of the effective dielectric constant in atomically thin materials, their 2D nature should be taken into account.\cite{keldysh_coulomb_1979} 

We also note that in a previous work Mott-VRH was observed in the transport behavior of n-butyllithium treated MoS$_2$ from two-terminal measurements, a hopping mechanism that is different compared to our analysis on data from four-terminal measurements.\cite{kim_electrical_2016} We note, however, that an analysis of our two-terminal resistance data from device A can also be performed with the Mott-VRH model, with an exponent of $a=0.31$ and a localization length of 0.7 nm, very similar to previous work,\cite{kim_electrical_2016} and in this sense the two datasets are not in contradiction.

Our observation of Efros-Shklovskii driven transport in 1T/1T$^{\prime}$-MoS$_2$ agrees with studies on similar systems.\cite{beloborodov_coulomb_2005}$^,$\cite{joung_coulomb_2011}$^,$\cite{joung_efros-shklovskii_2012} Theoretical studies of irregular arrays of metallic grains, embedded in an insulating matrix resembling the n-butyllithium treated MoS$_2$ lattice, show that Coulomb interactions take place and that the transport follows the ES-VRH mechanism.\cite{beloborodov_coulomb_2005} Similar results were obtained from electrical transport in two-dimensional graphene quantum dot arrays\cite{joung_coulomb_2011} and chemically reduced graphene oxide sheets.\cite{joung_efros-shklovskii_2012}

Finally, the significant increase of the contact resistance in 1T/1T$^{\prime}$-MoS$_2$ suggests that the material is not the ideal candidate for contacting semiconducting 2H-MoS$_2$ for experiments at cryogenic temperatures. Nevertheless, at room temperature, the measured $R_{c}$ is low, which is in agreement with previous studies.\cite{kappera_phase-engineered_2014} Measurements on devices with varying channel lengths (transfer length method) and studies on flakes with higher content of the metallic 1T phase, can provide more insights regarding the behavior of the current injection into the 1T/1T$^{\prime}$ material.	

\section{\label{sec:level1} Conclusion}
\noindent
In summary, we observe Efros-Shklovskii-VRH transport in 1T/1T$^{\prime}$-MoS$_2$, as obtained from a treatment with n-butyllithium. From temperature-dependent measurements in the Ohmic regime and electric-field dependent studies in the non-Ohmic and electric-field driven regime, we obtain localization lengths in the order 1-3 nm. An interesting future direction of research could be to quantify and control the mixing of the different phases and observe how this affects the transport mechanisms.\\

\begin{center}
  \textbf{\small ACKNOWLEDGMENTS}
\end{center}

\noindent
This work has been supported by the Organisation for Scientific Research (NWO) and the Ministry of Education, Culture, and Science (OCW). We thank Yaroslav M. Blanter, Holger R. Thierschmann and Dirk J. Groenendijk for fruitful discussions.

\bibliographystyle{apsrev4-1}
\bibliography{paper_for_arxiv}
\bigskip


\pagebreak
\clearpage 

\widetext
\setcounter{equation}{0}
\setcounter{figure}{0}
\setcounter{table}{0}
\setcounter{page}{1}
\makeatletter
\renewcommand{\theequation}{S\arabic{equation}}
\renewcommand{\thefigure}{S\arabic{figure}}
\renewcommand{\bibnumfmt}[1]{[S#1]}
\renewcommand{\citenumfont}[1]{S#1}


%
%
%
%
%
%
\end{document}